\documentclass[fleqn,11pt]{article}
\usepackage{hyperref}
\usepackage{amsmath, amssymb, multirow, graphicx,footnote,caption}

\usepackage{graphicx}
\usepackage{float,epsfig}
\usepackage{latexsym}
\usepackage{amsmath}
\usepackage{amssymb}
\usepackage{amsthm}
\usepackage{url}
\usepackage{subfigure}

\usepackage{changepage}

\makeatletter
\newcommand{\Rmnum}[1]{\expandafter\@slowromancap\romannumeral #1@}
\makeatother
\begin{document}\sloppy
	
	\vspace*{0.2in}
	
	\begin{flushleft}
		{\Large
			\textbf\newline{Robust modelling framework for short-term forecasting of global horizontal irradiance} 
		}
		\newline
		\\
		Edina Chandiwana\textsuperscript{1,2},
		Caston Sigauke\textsuperscript{1*},
		Alphonce Bere\textsuperscript{1}
		\\
		\bigskip
		\textbf{1} Department of Mathematical and Computational Sciences, University of Venda, Private Bag X5050, Thohoyandou 0950, Limpopo, South Africa.
		\\
		\textbf{2} Department of Applied Mathematics, Midlands State University, Private Bag 9055, Senga, Gweru.\\
		\bigskip
		
		%
		%


		\textcurrency Current Address: Department of Mathematical and Computational Sciences, University of Venda, Private Bag X5050, Thohoyandou 0950, Limpopo, South Africa.
	\end{flushleft}
	
	\section*{Abstract}
	The increasing demand for electricity and the need for clean energy sources have increased solar energy use. Accurate forecasts of solar energy are required for easy management of the grid. This paper compares the accuracy of two Gaussian Process Regression (GPR) models combined with Additive Quantile Regression (AQR) and Bayesian Structural Time Series (BSTS) models in the 2-day ahead forecasting of global horizontal irradiance using data from the University of Pretoria from July 2020 to August 2021. Four methods were adopted for variable selection, Lasso, ElasticNet, Boruta, and GBR (Gradient Boosting Regression). The variables selected using GBR were used because they produced the lowest MAE (Minimum Absolute Errors) value. A comparison of seven models GPR (Gaussian Process Regression), Two-layer DGPR (Two-layer Deep Gaussian Process Regression), bstslong (Bayesian Structural Time Series long), AQRA (Additive Quantile Regression Averaging), QRNN(Quantile Regression Neural Network), PLAQR(Partial Linear additive Quantile Regression), and Opera(Online Prediction by ExpRt Aggregation) was made. The evaluation metrics used to select the best model were the MAE (Mean Absolute Error) and RMSE (Root Mean Square Error). Further evaluations were done using proper scoring rules and Murphy diagrams. The best individual model was found to be the GPR. The best forecast combination was AQRA ((AQR Averaging) based on MAE. However, based on RMSE, GPNN was the best forecast combination method. Companies such as Eskom could use the methods adopted in this study to control and manage the power grid. The results will promote economic development and sustainability of energy resources. \\
	
	{\bf Key words:} Additive quantile regression; Bayesian structural time series; Forecast combination; Gaussian processes; Solar irradiance.
	
	\section{Introduction}
	Over the past years, South Africa has suffered an energy crisis due to climatic changes. This has resulted in power companies like Eskom resorting to alternative energy sources like solar power, also referred to as Global Horizontal Irradiance (GHI). Solar power has been preferred as an alternative energy source that is inexhaustible, sustainable, highly abundant, cheap and does not pollute the environment. This has led to solar power forecasting becoming an important aspect of the energy management system since solar power generation is directly linked to the management of the power grid Yang et al. ~\cite{Yang}.	The United Nations, Sustainable Development Goal (SDG) number 7, on affordable and clean energy, UNDP ~\cite{UNDP}, encourages the use of renewable energies to mitigate climatic changes, which makes the prediction of GHI vital. The goal is to promote the utilisation of renewable energy sources such as solar and wind because they are affordable, reliable and adequate.
	
	Various methods have been proposed to predict GHI generation at different scales. The current methods are statistical, machining learning, and numerical weather prediction models. The current study uses a 2-days-ahead probabilistic modelling framework to predict short-term solar power generation. A comparison is made between Bayesian Structural Time Series (BSTS), Deep Gaussian Regression (DGP), Gradient Boosting Regression (GBR) and the Gaussian Process Regression coupled with Quantile Regression.  	
	
	Several authors have done research applying Gaussian regression, ~\cite{Bilionis}, ~\cite{Tolba},~\cite{Wang}, and  ~\cite{Zhang} among others. Billionis et al. ~\cite{Bilionis} proposed
	a recursive Gaussian Process approach reduces the input space of satellite-based observations to perform iterated predictions. They first applied factor analysis for dimensionality reduction to come up with two maps, one for reconstruction and the other for reduction. Their results proved that the proposed method performed worse than the ground-based model. Tolba et al. ~\cite{Tolba} considered the prediction of GHI using Gaussian Process Regression(GPR). They applied GPR to forecast horizontal data, ranging from 30 minutes to 5 hours. Their main thrust was on the selection of the kernel functions. They used simple and complex kernels and compared their results to the persistent model. Their results showed that when the quasiperiodic kernel was applied, the GPR model outperformed the other GPR models and the persistent model.  One of the few studies that looked at GPR was done by Wang et al. ~\cite{Wang}. They proposed a hybrid model combining LSTM (Long Short-Term Memory) and Gaussian Process Regression. The method is compared to neural networks. The results showed that the proposed method produced superior results.
	
	Zhang et al. ~\cite{Zhang} predicted solar power using k means time series and Gaussian Process Regression. They clustered solar power radiation using different feature input categories. Different covariance functions of GPR were applied, and a comparative analysis was done. Kernel functions were selected using deterministic and probabilistic categories. The results of the proposed GPR proved to be superior to the Artificial Neural Network benchmark model used. Yu et al. ~\cite{Yu} developed a method of extracting trajectories called the Gaussian process factor analysis. They looked at spike trains that were first smoothed using a smoothing technique that accounted for spiking variability. Stonski ~\cite{Stonski} did a comparative analysis of feedforward layered neural networks on stochastic and Gaussian Process analyses. The results showed that the two had similar forecasting accuracy and better performance than the linear regression model. Al-Shedivat et al. ~\cite{Al-Shedivat} applied long, short-term memory recurrent networks coupled with GPs. Using expressive closed-form kernels, they used marginal likelihood estimation on a convergent semi-stochastic gradient method and exploited how kernels should be structured for forecasting. The models were applied to car applications which are self-driven, system identification, and power forecasting, and the results showed that the proposed method was efficient and convergent as required. Tsymbalov et al. ~\cite{Tsymbalov} applied a dropout-based model based on Bayesian Regression and Neural network coupled with Gaussian process on a chemical and physical real-life dataset. The results showed that the methods performed well for neural architectures involving dropout and Bayesian neural networks.
	
	Chandiwana et al. ~\cite{Chandiwana} did a study on predicting global horizontal irradiance (GHI) using Gaussian process regression. They predicted GHI based on GPR coupled with core vector regression. The performance of this method was compared with that of two benchmark models: support vector regression (SVR) and gradient boosting regression (GBR). The proposed methodology performed better than the benchmark models.
	
	Over the past few years, researchers have been exploring the use of machine learning techniques which produce accurate forecasts. Most GPR methods highlighted in the literature that has looked at predicting solar power have proved to produce accurate predictions, but these models need to be improved. Predicting solar power is difficult because of the uncertainty and variable nature of the datasets. To the best of our knowledge, little has been done to date on forecasting South African Global Horizontal Irradiance data using Gaussian Process Regression. Hence there is a need to explore this area. The technique used in this research, GPR coupled with quantile regression, addresses that aspect of dealing with uncertainty, and the combination of forecasts was done to improve forecasting accuracy. GPR was combined with additive QR (AQR), quantile regression neural network (QRNN), and partial linear additive quantile regression (PLAQR) models together with the use of a convex combination method.
	
	This research aims to develop a forecasting model that accurately predicts 2-day-ahead solar power. We compared two models, BSTS and GPR, to select the best among these two models. Finding solutions to uncertainty and variability is crucial in energy forecasting. We propose the GPR to solve this problem, combining it with quantile regression. Prediction models are being developed daily, but an inseparable part needs attention: error reduction and accurately capturing uncertainty. The bigger the error is, the poor the prediction outcomes will be. Thus it is crucial to reduce errors to improve the accuracy of results. We expect to develop an improved modelling framework for GHI by adopting the combined hybrid models, thus expecting a reduction of errors and, at the same time dealing with variability and uncertainty.
	
	\subsection{Research highlights}
	The first contribution of this research is the coupling of the Gaussian process with QR to develop a modelling framework for medium-term forecasting of GHI. This approach provides estimates that cater for the dataset's uncertainty and scalability, hence its strong performance.
	
	Secondly, we applied a modelling framework based on Bayesian Structural time series on GHI data. This model's performance is superior because it incorporates other variables rather than historical predictors and avoids over-fitting. BSTS avoids over-fitting and captures the correlation between many state components and multiple time series. This model is a combination of time series and Bayesian techniques.
	%
	
	Lastly, the combination of forecasts was done using GPNN, PLAQR, QRA and Opera. These were combined with BSTSlong and GPR. Bates and Granger ~\cite{Bates} concluded in their research that combining forecasts reduces errors.
	The rest of the paper is organised as follows: The first section describes the methods, followed by the section on empirical data analysis, a section representing the discussions and finally, the conclusion.
	
	\section{Materials and methods}
	\label{sec:Materials and Methods}
	\subsection{Schematic presentation of methodology}
	
\begin{figure}[H]
		\includegraphics[width=1.07\linewidth, height=0.60\textheight]{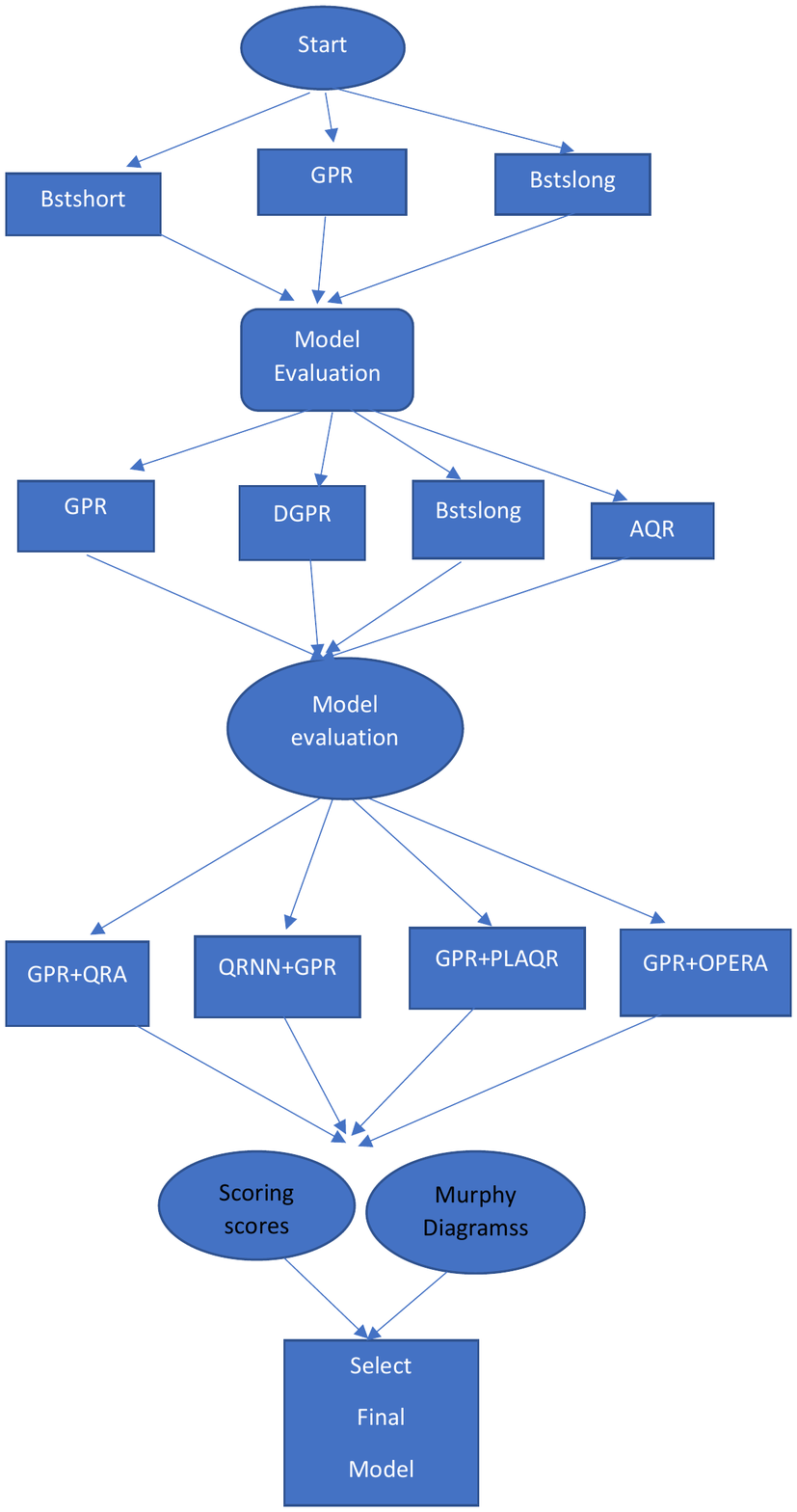}
		\vspace{-0.5in}
		\caption{\bf Methodology flow chart.}
		\label{fig:rplot01}
	\end{figure}

	Fig~\ref{fig:rplot01} shows the schematic presentation of the adopted methodology, with all the stages laid out step by step. The data is split into training and testing sets. The training data set is used in training the models to determine the most appropriate fitting model. This is followed by using the trained models to produce two-day ahead predictions. Evaluation metrics are then used to determine the model with the highest predictive capability. The forecasts produced by the unique models are combined using four robust forecast combination methods. Proper scoring rules and Murphy diagrams are then used to determine the forecast combination method that produces robust medium-term forecasts of GHI.
	
	\subsection{Models}
	\subsubsection{Gaussian process}	
	
	A Gaussian process (GP) was used in this research. It is a nonparametric probabilistic stochastic machine learning technique based on multivariate normal distribution kernels. A GP is based on a function with a continuous domain and defined by its mean $m(x)$ and covariance $k(x_i, x_j)$ where $x_i$ and $x_j$ are two points that influence each other. If we have a GP described by a function defined by $f(x(t)), x(t) \in \mathbb{R}^d$, where $x(t)$ are the weather variables like air temperature, wind speed and relative humidity, indexed over time $t$, $f(x)$ is the response of these weather variables into a $d$ dimensional space.
	
	A GP has a mean given in Eq~\ref{eq1},
	\begin{equation}E(f(x))=m(x)
		\label{eq1}
	\end{equation}
	and covariance function,
	
	\begin{multline}k(x_i,x_j)=E[\{f(x_i)-m(x_i)\}\{f(x_j^1)-m(x_j^1)\}].\\
		\label{eq2}
	\end{multline}
	
	Therefore, the function describing the GP of GHI is given in Eq~\ref{eq3}.
	\begin{equation}
		g(x)\sim GP(m(x), k(x,x^1))
		\label{eq3}
	\end{equation}	
	The method is named after the famous Carl Fredrich Gauss because it is based on the idea of the Gaussian distribution, commonly known as the normal distribution.
	
	Gaussian process regression (GPR) is a powerful tool used for regression and classification, and it performs better than other forecasting methods since it predicts uncertainty directly, giving reliable estimates. The forecasts from GPR are reliable because GP uses prior information, and different specifications of the covariance function are used.
	
	\subsubsection{Covariance functions}
	A GPR is defined by a mean and covariance function. A covariance function is a key component of the GPR model. There are various covariance functions: Matern, Linear, periodic, Radial basis function, Polynomial, Bessel including Spline. This study uses the radial basis function (RBF). The RBF is given in Eq~\ref{eq:RBF}
	\begin{equation}
		RBF=\sigma^2 \exp\left( \frac{(x_i-x_j)^2}{2l^2}\right) +\sigma_{ij}\sigma^2_{\mbox{noise}},
		\label{eq:RBF}
	\end{equation}
	where $l$ is the length parameter, $\sigma^2$ is the variance and $\sigma ^2_{noise}$ is the variance of the noise term.
	
	\subsection{Benchmark models}
	\subsection{Deep Gaussian process regression}
	
	The deep Gaussian process (DGP) is defined in Saucer et al. ~\cite{Saucer} as a hierarchical layering of GPs. Each layer is assumed to follow a multivariate normal distribution (MVN).
	
\begin{figure}[H]
		\includegraphics[width=0.9\linewidth, height=0.35\textheight]{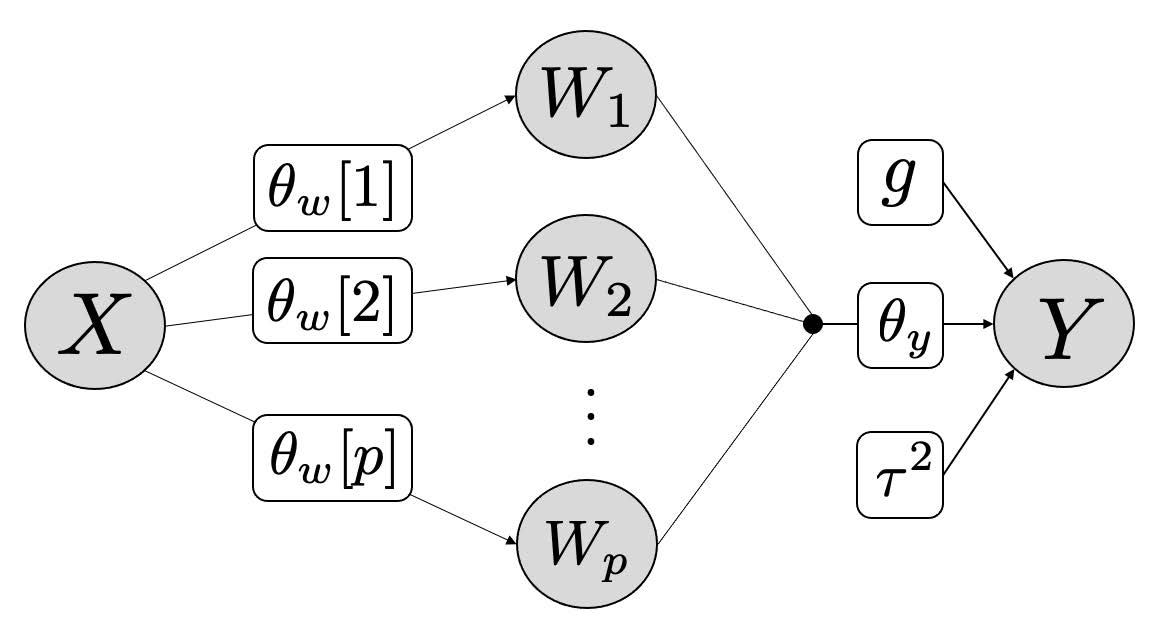}
		\caption{Two-layer deep GP model with $p$ latent nodes. Source: Saucer \cite{Saucer}.
			\label{layer}}
	\end{figure}

	Let $X_n$ represent $n \times d$ inputs with $Y_n = f(X_n)$ denoting the functional evaluations. That is $X_n \rightarrow Y_n$,  $f: \mathbb{R}^d \rightarrow \mathbb{R}$ and $Y_n\thicksim N_n(\mu,\Sigma)$. See Saucer \cite{Saucer} for details.   For the inputs, $X_n$, to reach the response, $Y_n$ has to pass through several intermediate GPs. In this study, we shall consider the two-layer model presented in Figure \ref{layer}.
	
	In Figure \ref{layer}, $W$ denotes the hidden GPs (latent nodes), and the structure is described hierarchically as Saucer (\cite{Saucer}):
	\begin{eqnarray*}
		Y_n|W\thicksim N_n(0,\Sigma(W)) \\
		W\thicksim N_n(0,\Sigma(X_n))
	\end{eqnarray*}
	The present study will consider two layers only since increasing the number of layers will not significantly improve the results but will only result in high computational cost (Saucer, \cite{Saucer}, Radaideh, \cite{Radaideh}). Estimating the parameters is through a hybrid Gibbs-ESS-Metropolis algorithm discussed in detail in Saucer, \cite{Saucer}.
	
	\subsection{Bayesian structural time series}
	Bayesian structural time series (BSTS) was also used in this research. It is a machine learning technique
	used in time series for forecasting, Brodersen, Scott, \cite{Brodersen, Scott}. A BSTS model is constructed as a state-space model, and there are two
	pieces to the model. The first component is the observation equation given in Eq~\ref{eq5}.
	\begin{equation}Y_{it}=\beta^Tx_{it}+Z_t^T\alpha_t+\varepsilon_{it}
		\label{eq5}
	\end{equation}
	The second component of the BSTS model is a transition equation that defines how the latent states evolve.
	\begin{equation}\alpha_{a+1}=T_t\alpha_t+R_t\eta_t,
		\label{eq6}
	\end{equation}
	where $Y_{it}$ is GHI, $t$ represents time, $x_{it}$ are the meteorological variables at time $t$, $\eta_t\sim N(0,\sigma^2_{\eta})$, $\alpha_t$, $Z_t, T_t$ and $R_t$ are structural parameters and $\beta^{T}X_{it}$ is the regression component.
	
	BSTS introduces a regression aspect developed from the structural model's independent variables. It deals with many independent variables by coming up with latent factors. It induces sparsity on the coefficients, putting a spike and slab prior to distribution. The model uses three components, spike and slab, Kalman filter and Bayesian model averaging.
	
	\subsubsection{Spike and slab}
	Spike and slab is a Bayesian variable selection technique. The idea was first developed by Mitchel ~\cite{Mitchell}. It was further developed by Madigan ~\cite{Madigan}. This Bayesian variable selection approach is made by mixing priors using a spike and slab for the effects subject to selection. Variable selection is a major step in regression analysis. The model parameters of the regression model $p(\mu, \sigma^2, \alpha)$ are specified with priors having the structure in Eq~\ref{eq:priors}.
	\begin{equation}p(\mu, \sigma^2, \alpha)=p(\mu, \sigma^2).p(\alpha \mid \sigma^2,\mu)
		\label{eq:priors}
	\end{equation}		
	with the probability of mean and error variance being $p(\mu,	\sigma^2)=\frac{1}{\sigma^2}$. The spike and slab priors are given in Eq~\ref{eq:prior2}.
	\begin{equation}
		p(\alpha \mid \delta) =p_{slab}(\alpha_{j}).\prod_{j:\delta_{j}=0}p_{spike}\alpha_{j},
		\label{eq:prior2}
	\end{equation}
	where $\delta_{j}$ represents an indicator variable and $\alpha_{j}$ is the regression effect.
	
	\subsubsection{Kalman filter}
	The time series feature is solved using a Kalman filter, Harvey ~\cite{Harvey}, Durbin and Koopman ~\cite{Durbin}. A  Kalman filter can be defined as an equation that gives a recursive solution to the least-squares method. It is an optimal iterative estimation algorithm whose objective is to minimise error. A Kalman filter is given in Eq~\ref{kalman}
	
	\begin{equation}K=P_{j}C^{T}(CP_{j}C^{T}+R_{t})^{-1}
		\label{kalman}
	\end{equation}
	Eq.\ref{gain} is called the Kalman gain.
	
	\begin{equation}P_{j+1}=P_{j}-KCP_{j},
		\label{gain}
	\end{equation}
	\newline where P is the state error, C is the observational model, and R is the noise covariance matrix.
	
	\subsubsection{Bayesian model averaging}
	The Bayesian model averaging (BMA), developed by ~\cite{hoeting}, is a parameter estimation method obtained from averages of the predictions. BMA is a method that analysts prefer because it reduces underestimation of uncertainty, produces optimal forecasts under several loss functions, and is robust to model misspecification.
	
	The average posterior distribution of each model is given in Eq.~\ref{eq9}
	\begin{equation}p(GHI\mid D)= \sum_{j=1}^{J} p(GHI\mid A_J,D) \times P(A_j \mid D),	
		\label{eq9}
	\end{equation}
	where $A_1,...,A_J$ are the models considered. The posterior probability of the model $A_J$ is given in Eq.~\ref{eq10}
	\begin{equation}
		\displaystyle	p(A_J\mid D)=  \frac{p(D\mid A_{l})\times p(A_{J})}{\sum_{l=1}^{J} p(D\mid A_J)\times P(A_l)},
		\label{eq10}
	\end{equation}
	where the likelihood of $A_j$ is given in Eq~\ref{eq11}
	\begin{equation}p(D\mid A_J)=\int p(D\mid \theta_j, M_j) \times p(\theta_j\mid M_j)d\theta_j,
		\label{eq11}
	\end{equation}
	where $ \theta_j$ is a vector parameter of model $A_J$.
	
	\subsubsection{Gradient boosting regression}
	Gradient Boosting regression is one of the Benchmark models used in this research. It is a machine learning method used in regression analysis and classification tasks. It is used for building forecasting models by using decision trees to generate negative gradients using samples and introducing a weak learner to relate to the available weak learners. The function $F(x)$, which approximates the best values of the predicted output variables, is given by
	\begin{equation}
		\hat{F}=arg_{F}  \hspace{0.1cm} min E_{x,y}\left[ L(y,F(x))\right]
		\label{loss}
	\end{equation}	
	where $L(y,F(x))$ is the loss function.
	GBR aims at minimising the loss function in 	\ref{loss}.
	\begin{equation}
		L(y,F(x))=(Y,F(x))^2
	\end{equation}
	we minimise the function in \ref{loss} by adjusting $F(x_1), F(x_2),F(x_3),...,F(x_n)$.
	
	Table~\ref{tab:t1} presents a summary of some of the advantages and disadvantages of the proposed models, which are GPR, two-layer DGPR and BSTS.
	\begin{table}[H]
		\small
		\centering
		\caption{Model comparisons.}
		\label{tab:t1}
			\begin{tabular}{|l|l|l|} 	\hline
				Models            & {\bf Strengths}                             &  {\bf Weaknesses}     \\ \hline
				GPR (M1)	      & 1. It is robust and versatile.                         & 1. It is computationally expensive.\\
				& 2. It accurately captures model uncertainty.   & 2. They are not sparse.\\
				& 3. It offers a probabilistic approach        &      \\
				& convenient for solving stochastic problems.  &      \\
				& 4. Allows addition of prior knowledge        &    \\
				& and specification.                           &   \\
				& 5. Allows scalability of large datasets.       &   \\
				& 6. Tune hyper-parameters by maximising &   \\
				& the marginal likelihood.             			&   \\ \hline
				BSTS (M2)     & 1. Handles uncertainty well.     & 1. Computational analysis is extremely   \\
				&                                 & difficult for the posterior distribution. \\
				& 2. It can handle a large number of variables.         &   \\
				& 3. Can predict stochastic patterns of  			&   \\
				& the time series accurately.                       &        \\
				& 4. Parameters of the model change over time.      &  \\
				& 5. Uses prior information on the variables.  &  \\  \hline
				Two-layer DGP (M3)  &  1. Nonstationary flexibility.  & 1. Computational cost is high.   \\
				& 2. Ability to cope with abrupt  &  2. MLE point estimtes lead to        \\	
				& regime changes in training data.    & overfitting, treating noise       \\
				& 3. Nonlinear predictive capability.    &  as a signal.         \\  \hline	
				GBR(M4)&1. It is flexible and sensitive to outliers&1. The model may cause\\
				&2. No data pre-processing required&overfitting\\
				&&\\
				&3. Conversion of weak learners into strong learners& \\ \hline
			\end{tabular}
	\end{table}
	
	\subsection{Forecasts combination}
	
	\subsubsection{Additive quantile regression model}
	The study used the quantile generalised additive model (quantGAM) based on the work of~\cite{gaillard2016} and extended by~\cite{fasiolo2020}, defined as:
	\begin{equation}
		y_{t,\tau}=\sum_{j=1}^ps_{j,\tau}(x_{tj})+\varepsilon_{t,\tau};~~\tau\in(0,1).
		\label{eq6}
	\end{equation}
	
	The smoothing function, $s$, is written as:
	\begin{equation}
		s_{j\tau}(x)=\sum_{k=1}^q\beta_{kj}b_{kj}(x_{tj}),
		\label{eq7}
	\end{equation}
	where $\beta_{kj}$ denotes the $j${th} parameter and $b_{kj}(x)$ represents the $j${th} basis function with the dimension of the basis being denoted by $q$. The parameter estimates of Equation~\eqref{eq6} are obtained by minimizing the function given in Equation~\eqref{eq8} as:
	\begin{equation}
		q_{Y\vert X}(\tau) = \sum_{t=1}^n\rho_\tau\bigg(y_{t,\tau}-\sum_{j=1}^ps_{j,\tau}(x_{tj})\bigg),
		\label{eq8}
	\end{equation}
	where $q_{Y\vert X}(\tau)$ is the conditional quantile function of $\tau \in (0,1)$ and $\rho_{\tau}(u)=u[\tau-{\bf I}(u<0)]$ is the pinball function. 	In this study, we are interested in estimating extreme conditional quantiles, i.e., to combine forecasts, we shall call it quantile regression averaging (QRA).	
	
	\subsubsection{Quantile regression neural network}
	A Quantile regression neural network(QRNN) is a model composed of quantile regression coupled with neural networks. A QRNN is given in Eq. \ref{eqqrnn}
	
	\begin{equation}
		\begin{aligned}\min_{v,w}{} &\sum\theta|y_t-f(x_t-v-w)\\& +\sum(1-\theta)|y_t-f(x_t-v-w) \\& +\lambda_1\sum w^2_{ji}+\lambda^2\sum_{i}v_i^2, \\  \end{aligned}
		\label{eqqrnn}
	\end{equation}
	where $\lambda_1$ and $\lambda_2$ are regularisation parameters and
	\begin{equation}	
		f(x_t-v-w)=g_2 \left( \sum^m_{j=0}v_jg_1\left (\sum_{i=0}^{n}W_{ji}x_{it} \right ) \right).
		\label{eq14}
	\end{equation}
	Eq. \ref{eq14} is the neural network model where $g_1(.)$ and $g_2(.)$ are activation functions, which are sigmoid and linear, respectively. The weight parameters are $v$ and $w$, $m$ is the number of hidden layers, and $n$ is the number of inputs. Theta ($\theta$) is the $\theta^{th}$ Quantile, and $x_{it}$ are independent variables which are the weather variables, and $y_t$ is the GHI.
	
	\subsubsection{Partially linear additive quantile regression}
	The two techniques, Generalized Additive Model (GAM) and QR(Quantile Regression) models are combined to come up with the partially linear additive quantile regression (PLAQR). A GAM is a generalised linear model where the dependent variable GHI depends on a smooth function of the independent variable. GAMS was developed to come up with a combination of the properties of the generalised linear models with additive models. The GAM model is given in Eq. ~\ref{eq15}
	\begin{equation}
		g(E(Y))=\beta_0+s_1(x_1)+s_2(x_2)+...+s_m(x_m),
		\label{eq15}
	\end{equation}
	where $Y$ represents GHI which is the response variable and, $x_i$ denotes the covariates, $s_j$ are functions of the independent variable. These functions might differ; for example, $s_1(x_1)$ might be a factor model,  $s_2(x_2)$ a weighted mean, etc.
	
	\subsubsection{Online prediction by expert aggregation}
	Lastly, Online Prediction by ExpRt Aggregation (OPERA) was one of the methods used for forecasts combination and was developed by \cite{hoeting}. Given a set of observed values for a particular variable $Y$, GHI, with a sequence of values $y_1,y_2,...,y_n$, are the predicted values. For a given time step $t=1,2,...,n$, there are predictions from independent variables, which are the weather variables given by $x_{k,t}$, where $k$ is a set of finite methods $k=1,..., K$ combines a multiple of algorithms of online learning, and they predict forecasts. The OPERA forecasts are given by Eq.~\ref{eq16}.
	
	\begin{equation}\hat{y}_t=s\sum _{k=1}^K P_{k,t}x_{k,t}
		\label{eq16}
	\end{equation}
	
	Table~\ref{tab:t2} presents a summary of some of the advantages and disadvantages of the four forecast combination models used in this paper.
	\begin{table}[H]
		\centering
		\caption{Model comparisons of forecast combination models.}
		\label{tab:t2}
		\begin{adjustwidth}{0in}{0in}
			\begin{tabular}{|l|l|l|} 	\hline	
				Models        & {\bf Strengths}                     &  {\bf Weaknesses}     \\ \hline
				AQR 	      & 1. Estimates are more robust to     & 1. Parameters are harder to estimate \\
				&     non-normal errors.                            & than other regression techniques.  \\
				& 2. Estimates are robust against outliers.         & \\
				& 3. Characterisation of the data is richer.   &      \\
				& 4. It is invariant to monotonic transformations.     &    \\ 		\hline
				PLAQR        & 1.  It is flexible to model data that is complex.       &1. Challenge of the existence  \\
				&  2. It is more parsimonious than   & of non-smoothness of the loss function. \\
				& 	 other regression models.  				     &   \\
				&   3. Very stable and flexible for       &  \\
				& large-scale interpolation.               &    \\
				&                        &  \\  \hline
				QRNN        & 1. It is flexible to model data on      & 1. Not very well developed \\
				& 	modelling dependencies.  		     & for tasks that are not classified.  \\
				& 2. Scalability is relatively good      & 2. Sometimes gives overconfidence in  \\
				& 	with large datasets.   				 & predictions when applying training sample.   \\ 			\hline
				OPERA       & 1. Combines multiple individual techniques         &  1. Finding feature importance  \\
				&   to build a more powerful technique.               & is a challenge. \\
				& 2. Better prediction.           & \\
				& 3. Improved results as compared        & \\
				&  to individual models.                  &   \\
				&                                   &   \\ 			\hline
			\end{tabular}
		\end{adjustwidth}
	\end{table}
	
	\subsection{Data and variables}
	This research made use of hourly data from SAURAN (Southern African Universities Radiometric Network) website \url{https://sauran.ac.za/}). The data are from a radiometric station based at the University of Pretoria, South Africa, with Latitude:-25.75308° (E), Longitude: 28.22859 ° (S) and Elevation: 1410m and for the period July 2020 to August 2021, giving us a total of 8522 observations. The data was split into two, the training set and the test set, the ratio used for the training was 80:20.
	
	The response variable is GHI(Global Horizontal Irradiance), and the explanatory variables with their specific abbreviations and measuring units are given in Table~\ref{tab:t3}.
	\begin{table}[H]
			\caption{Independent variables: UPR Station.}	
			\centering
			\label{tab:t3}
			\begin{tabular}{|l|l|l|} 	\hline
				Name & Description & Measuring units\\ 	\hline
				Air Temperature    & Temp & $^oC$\\
				Relative Humidity & RH & \%\\
				Wind Speed & WS & m\slash s\\
				Barometer Pressure & BP & mbar\\
				Wind Direction & WD & $^o$\\
				Wind Direction Standard Deviation & WD\_Stv & $^o$\\
				Rain Total & Rain\_Tot & mm\\
				Maximum wind speed & WS\_Max & m\slash s\\ 	\hline
			\end{tabular}
	\end{table}
	
\begin{figure}[H]
		\centering
		\includegraphics[width=0.7\linewidth, height=0.4\textheight]{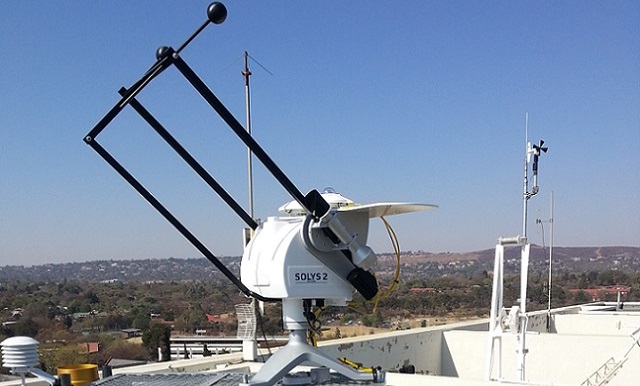}
		\vspace{0.3in}
		\caption{\bf UPR Weather instruments. Source \url{https://sauran.ac.za/}.}
		\label{fig:pyranometer}
	\end{figure}

	The SAURAN UPR station is located at the University of Pretoria, South Africa. The pyranometer is on top of a science building, giving good solar exposure. Figure \ref{fig:pyranometer} shows the position of weather instruments at the top of a building.
	
	\subsection{Evaluation metrics}
	The models' performances were assessed using RMSE (Root Mean Square Error) and MAE (Mean Absolute Error). Further evaluations were done using probability evaluation metrics, proper scoring rules, and Murphy diagrams.
	
	\subsubsection{Mean absolute error}
	The mean absolute error (MAE)  measures the average of all magnitude of errors. MAE is a measure of the accuracy of a given model. Absolute error is a measure of the amount of error in a model. It is defined in Eq. \ref{eq17}.
	\begin{equation}
		\mbox{MAE} = \frac{1}{m}\sum_{j=1}^{m}|y_i-\hat{y}_j|
		\label{eq17}
	\end{equation}
	
	\subsubsection{Root mean square error}
	Root Mean Square Error(RMSE) measures a quadratic scoring rule for the average magnitude of errors. It is defined in Eq.\ref{eq18}
	\begin{equation}
		\mbox{RMSE} = \sqrt{\frac{1}{m}\sum({y_i-\hat{y}_j}})^2,
		\label{eq18}
	\end{equation}
	where $y_i$ are the observed values, $\hat{y}_j$ are the predicted values and $m$ denotes the number of observations in the testing set.
	
	\subsection{Proper scoring rules}
	The proper scoring rules that were used in this research are logarithmic score (LogS), continuous ranked probability score (CRPS), and Dawid-Sebastian score (DSS). The 'ScoringRules' R package developed by Jordan, \cite{jordan} is used to compare and evaluate the methods used. A scoring rule assigns a penalty score, given by a function $S(F, y)$, which is used to evaluate a distribution $F$ over outcomes $y\in \mathbb{R}$ observed values. The scoring rules were used to assess the models. The lower the score, the better the forecasting accuracy.
	
	\subsubsection{Continuous ranked probability score}
	The continuous ranked probability score (CRPS) is used to evaluate probabilistic systems. It generalises the MAE to a probabilistic scenario. The score is calculated using Eq.~\ref{eq19}
	\begin{equation}
		\mbox{CPRS}(F,x) = \int_{-\infty}^{\infty} \left( F(y)-\mathbf{1}(y-x)\right)^2dy ,
		\label{eq19}
	\end{equation}
	where $\mathbf{1}(.)$ is the indicator function, it is a one if the function is positive or zero and 0; otherwise, $F(y)$ is the cumulative distribution function of $X$, which are weather variables, $y$ is the response variable, and $x$ are the independent variables.
	
	\subsubsection{Logarithmic score}
	It is a measurement that assesses how well the given variables perform on a certain dataset. The score is given in Eq~\ref{eq20}.
	\begin{equation}
		\mbox{LogS}(F,y)= -\mbox{log} f_{y},	
		\label{eq20}
	\end{equation}
	where $ f_{y}$ is the probability function and $y$ are the observed values.
	
	\subsubsection{Dawid-Sebastian score}
	It is a measure for evaluating the accuracy of multivariate forecasts. The score is given in Eq. \ref{eq21}
	\begin{equation}
		\mbox{DSS}(F,y)=\left(\frac{y-\mu_{F}}{\sigma_{F}}\right)^2 + 2\mbox{log}\sigma_{F},
		\label{eq21}
	\end{equation}
	where $y-\mu_{F}$ is the square error score, $\mu_{F}$ and $\sigma_{F}$ are the mean and variance, respectively.
	
	\subsubsection{Pinball Loss}
	The pinball loss function measures the accuracy of a quantile forecast. It is given by Eq.~\ref{eq25}.
	\begin{equation}
		\mbox{PL}(q\tau,t)= \left\{ \begin{array}{cl}2(1-\tau)|y_t-q_{\tau,t}|&if  \hspace{0.5cm}  y_t<q_{\tau,t}\\
			2\tau|y_t-q_{\tau,t} & if\hspace{0.5cm}  y_t 	\geq q_{\tau,t} \end{array} \right.,
		\label{eq25}
	\end{equation}
	where $y$ is the observed values, $q$-quantile forecasts at the $\tau^{\mbox{th}}$-is quantile.
	
	\section{Results} \label{sec:Results}
	
	\subsection{Exploratory data analysis}
\begin{figure}[H]
		\includegraphics[width=0.98\linewidth, height=0.35\textheight]{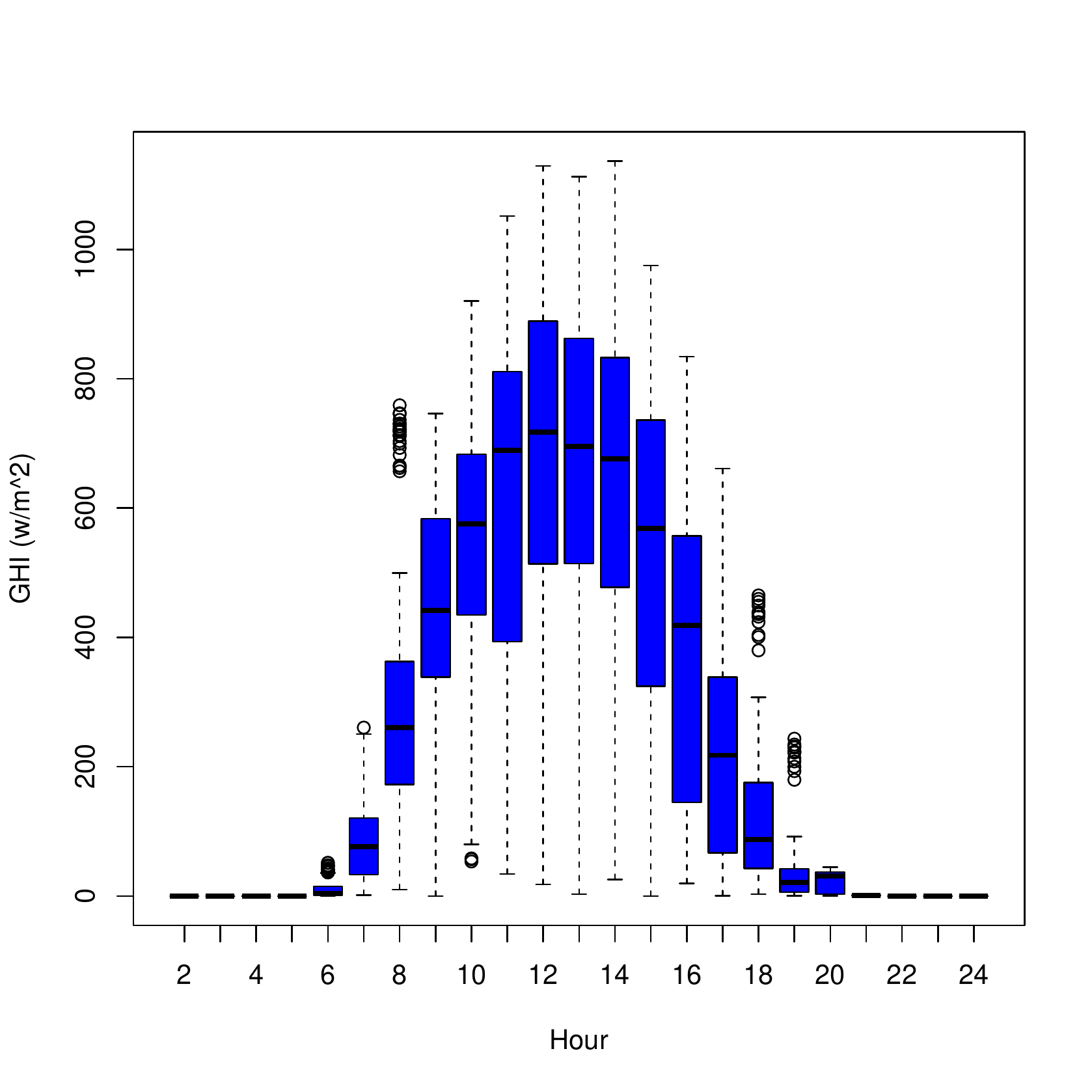}
		\includegraphics[width=0.98\linewidth, height=0.4\textheight]{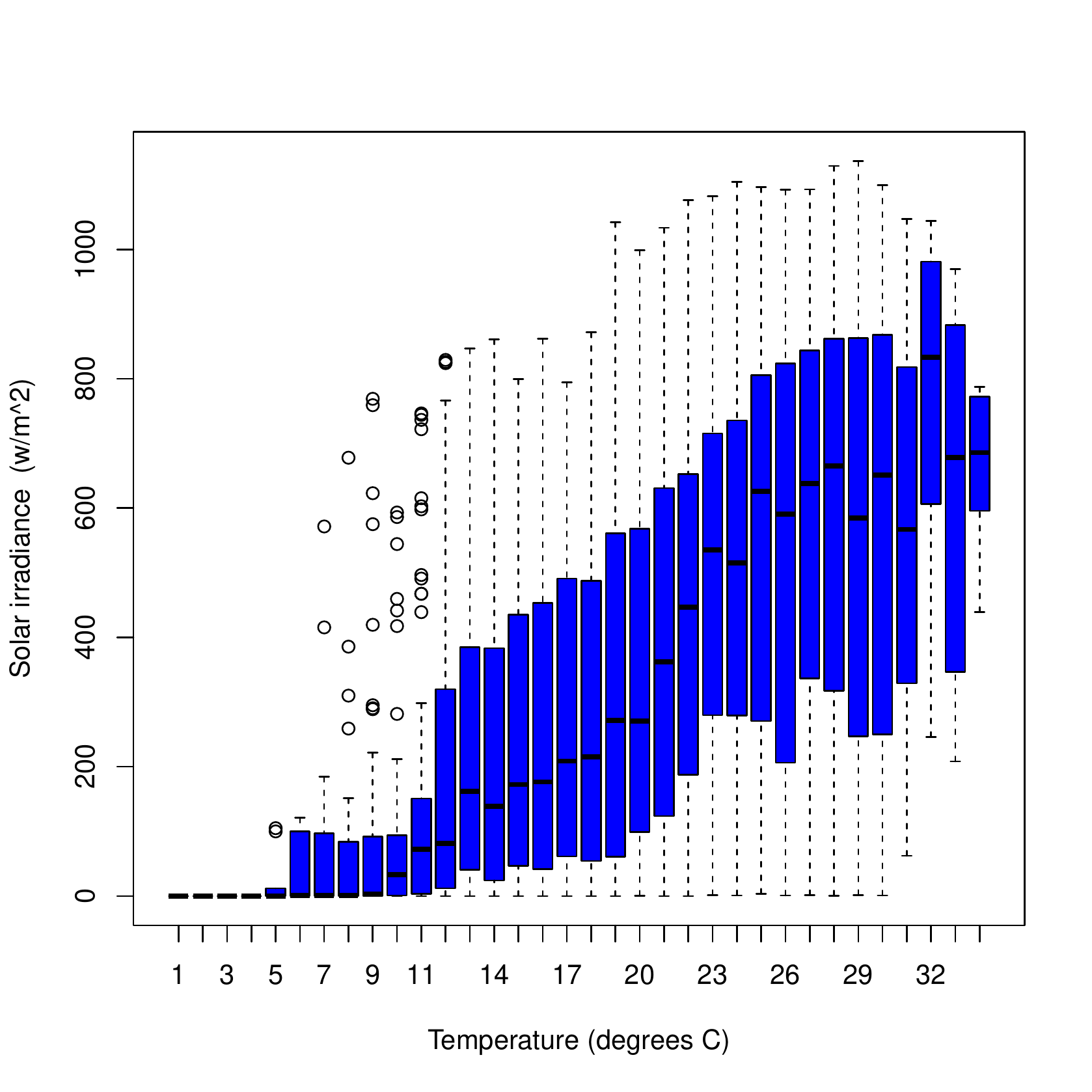}
		\caption{{\bf Top panel:} Distribution of hourly GHI. {\bf bottom panel:}  Joint distribution of GHI and temperature.}
		\label{fig:dist}
	\end{figure}

	%
	%
	The top panel of figure \ref{fig:dist} shows the distribution of hourly solar irradiance data. It shows that the sunlight hours are from 07:00 to 20:00. The largest amount of GHI is harvested between 11:00 and 16:00. The bottom panel, Figure \ref{fig:dist}, shows a plot of the joint distribution of GHI and temperature. As temperature increases, GHI also increases and seems to reach a peak around 29$^o$C then drops.
	
\begin{figure}[H]
		\includegraphics[width=0.9\linewidth, height=0.6\textheight]{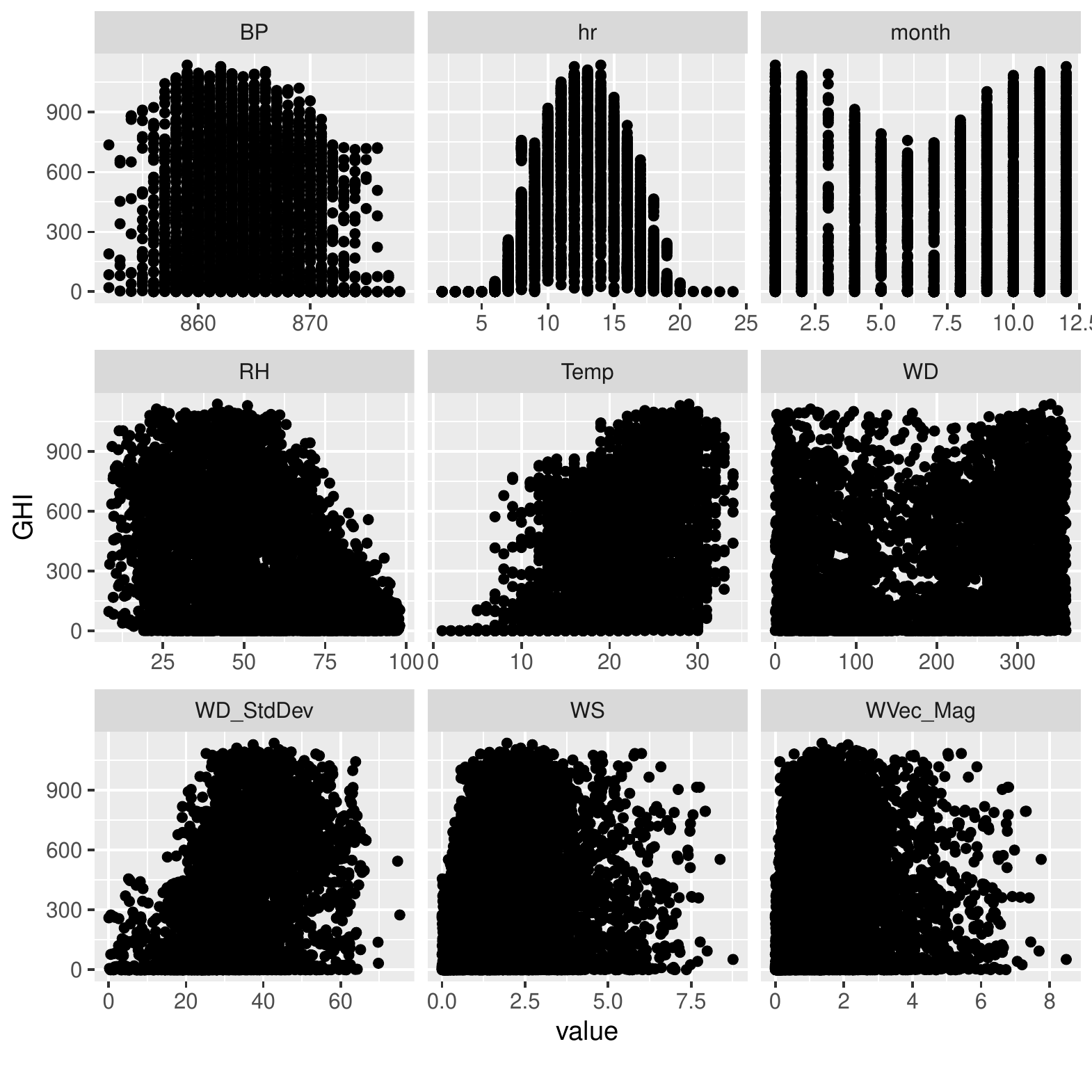}
		\caption{\bf Scatter plots of covariates against GHI.}
		\label{fig:plots}
	\end{figure}

	\newpage
	Scatter plots of the response variable GHI against covariates are given in Fig.~\ref{fig:plots}.  
	
	\subsection{\normalsize  \textbf{Variable Selection}}
	Variable selection is one of the most important challenges in statistical analysis. We considered five modern variable selection techniques: Lasso, ElasticNet, Boruta, GBM (Gradient Boosting Method) and MARS (multivariate adaptive regression splines). GBM (Gradient Boosting Method) was selected for variable selection because it had the lowest MAE compared to the other methods.
	
	A training dataset was used for variable selection, and then Linear Quantile Regression (LQR) was applied to the different methods. The evaluation results from the LQR models concerning the variable selection method are shown in Table~\ref{tab:t4}.
	
	\begin{table}[H]
		\centering
			\caption{Variable Selection.}	
			\begin{tabular}{|l|c|}			\hline
				Method & MAE\\ 			    \hline
				Lasso    & 100.48\\
				ElasticNet & 105.43\\
				MARS & 111.54\\
				BORUTA & 100.55\\
				GBR & 100.36\\ 			\hline
			\end{tabular}
			\label{tab:t4}
	\end{table}
	
\begin{figure}[H]
		\includegraphics[width=1.0\linewidth, height=0.3\textheight]{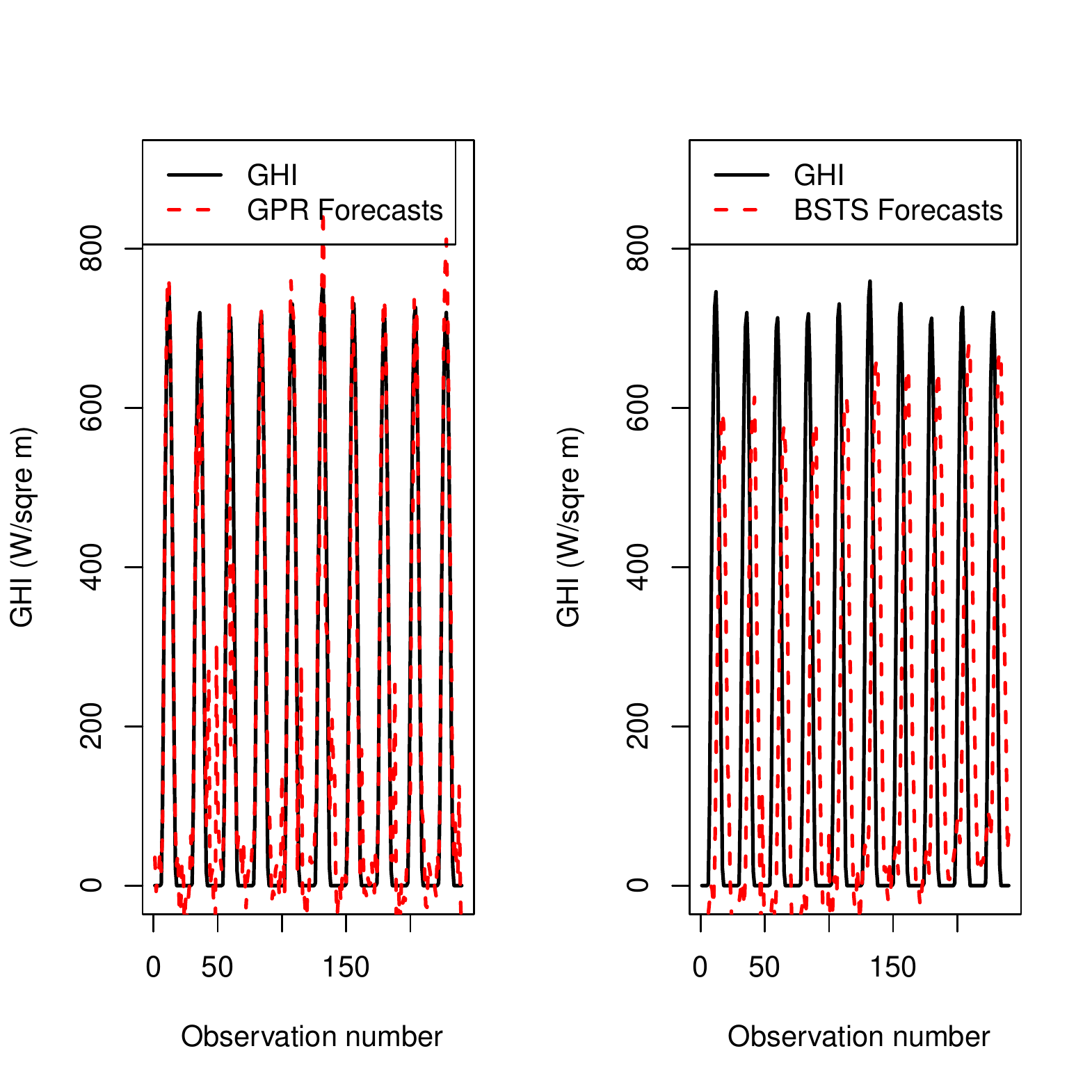}  \\
		\includegraphics[width=1.0\linewidth, height=0.3\textheight]{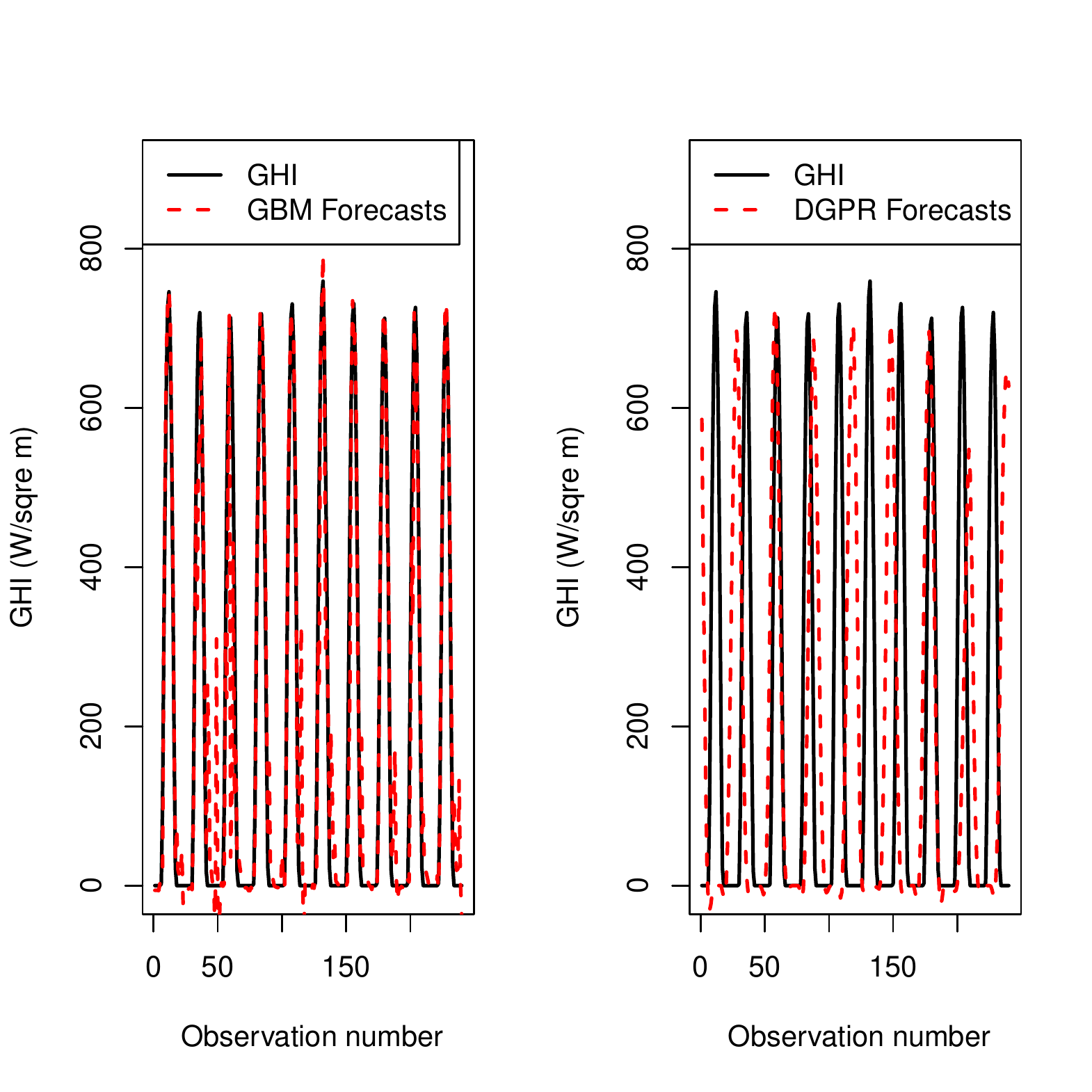}
		\includegraphics[width=1.0\linewidth, height=0.3\textheight]{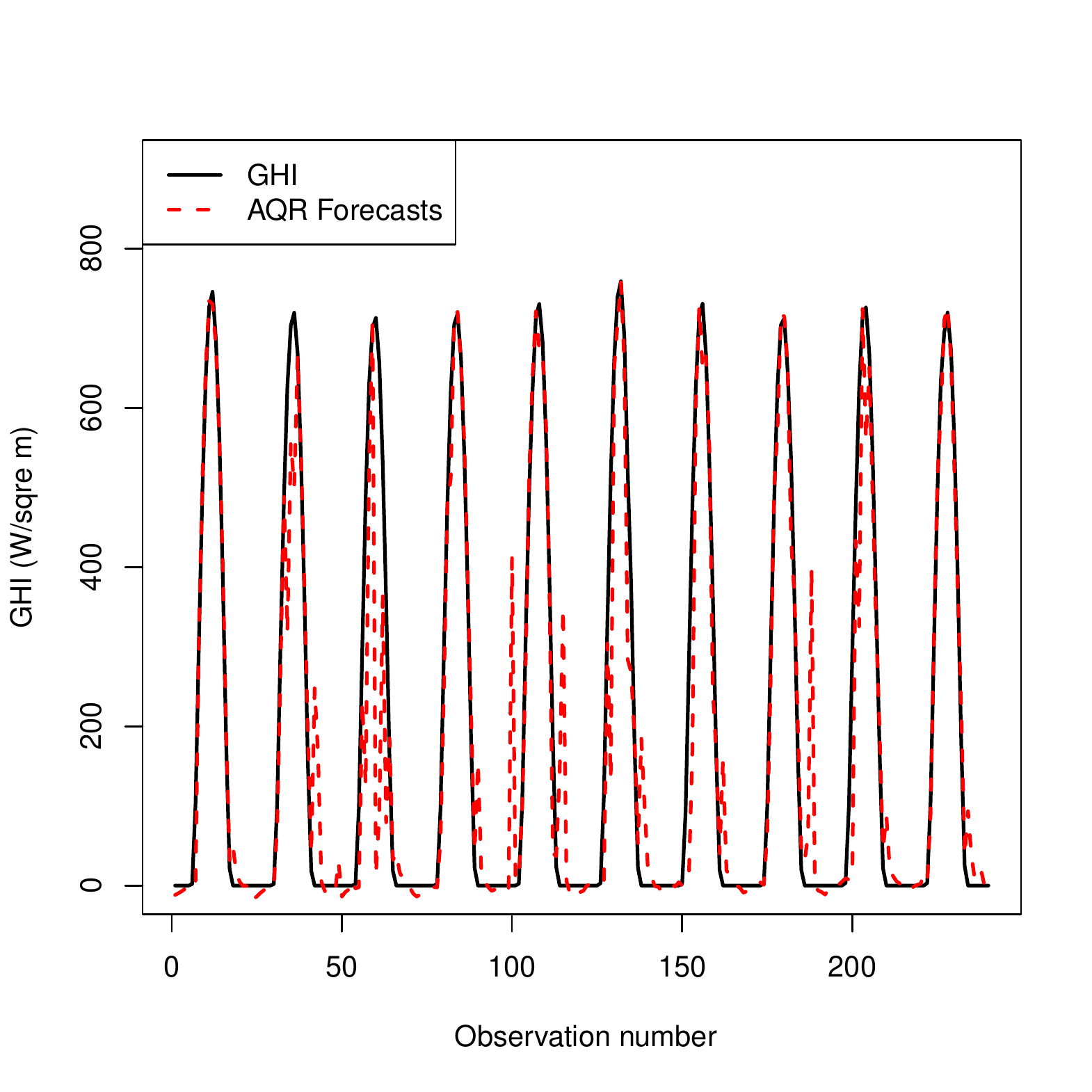}
		\caption{\bf GHI vs predicted using the developed models.}
		\label{fig:all}
	\end{figure}

	The models were then used for two-day ahead forecasting of GHI. The two-day ahead forecasts superimposed with observed GHI are given in Fig~\ref{fig:all}. The plots show a smooth fit of the predicted in red and GHI in black.
	
	A comparative analysis of the GPR model was done with that of BSTS short and long prediction, GBM, AQR and Two-layer DGPR models and based on MAE and RSME as the evaluation metrics. The comparison results are shown in Table~\ref{tab:t5}. The GPR model proved to be more efficient than the other models since it had low values of RMSE and MAE. We decided that the BSTS\_short and Two-layer DGPR were not appropriate models since they produced very high values of MAE and RMSE. The next section is a combination of forecasts, where we will combine the GPR with other quantile regression models and the OPERA model.
	
	\begin{table}[H]
		\centering
			\caption{Comparison of the models.}
			\label{tab:t5}
			\begin{tabular}{|l|l|l|}			\hline
				Model & RMSE & MAE\\ 		    \hline
				GPR   & 75.92 & 45.32\\
				BSTS\_short & 3010.72 & 2533.07\\
				BSTS\_long & 257.482 & 196.14\\ 	
				GBM        & 78.37  & 36.26 \\
				AQR        & 92.83  & 37.49  \\
				Two-layer DGPR & 360.74 & 256.55 \\		\hline
			\end{tabular}
	\end{table}
	
	\subsubsection{Combining forecasts}
	As a way of improving the accuracy of forecasts, forecasts were combined. Bates and Granger \cite{Bates} found that combining forecasts enhances the accuracy of forecasts. We used forecasts from the following models: quantile regression averaging (QRA), QRNN, PLAQR and OPERA, a convex combination method. The first method, linear Quantile Regression, was combined with bstslong and GPR. We are going to use fQRA to represent the produced forecasts. The second method used is Quantile regression neural network combined with bstslong and GPR. The combined forecasts will be referred to as the QRNN model. The third method is Partially linear Additive Quantile Regression combined with bstslong and GPR. The combined forecasts will be referred to as PLAQR. The fourth method uses the convex combination using the OPERA R package. The combined forecasts will be referred to as OPERA. Table~\ref{tab:t6} shows the results for forecasting the accuracy of the four methods based on two evaluation metrics.
	
\begin{figure}[H]
		\includegraphics[width=1.0\linewidth, height=0.4\textheight]{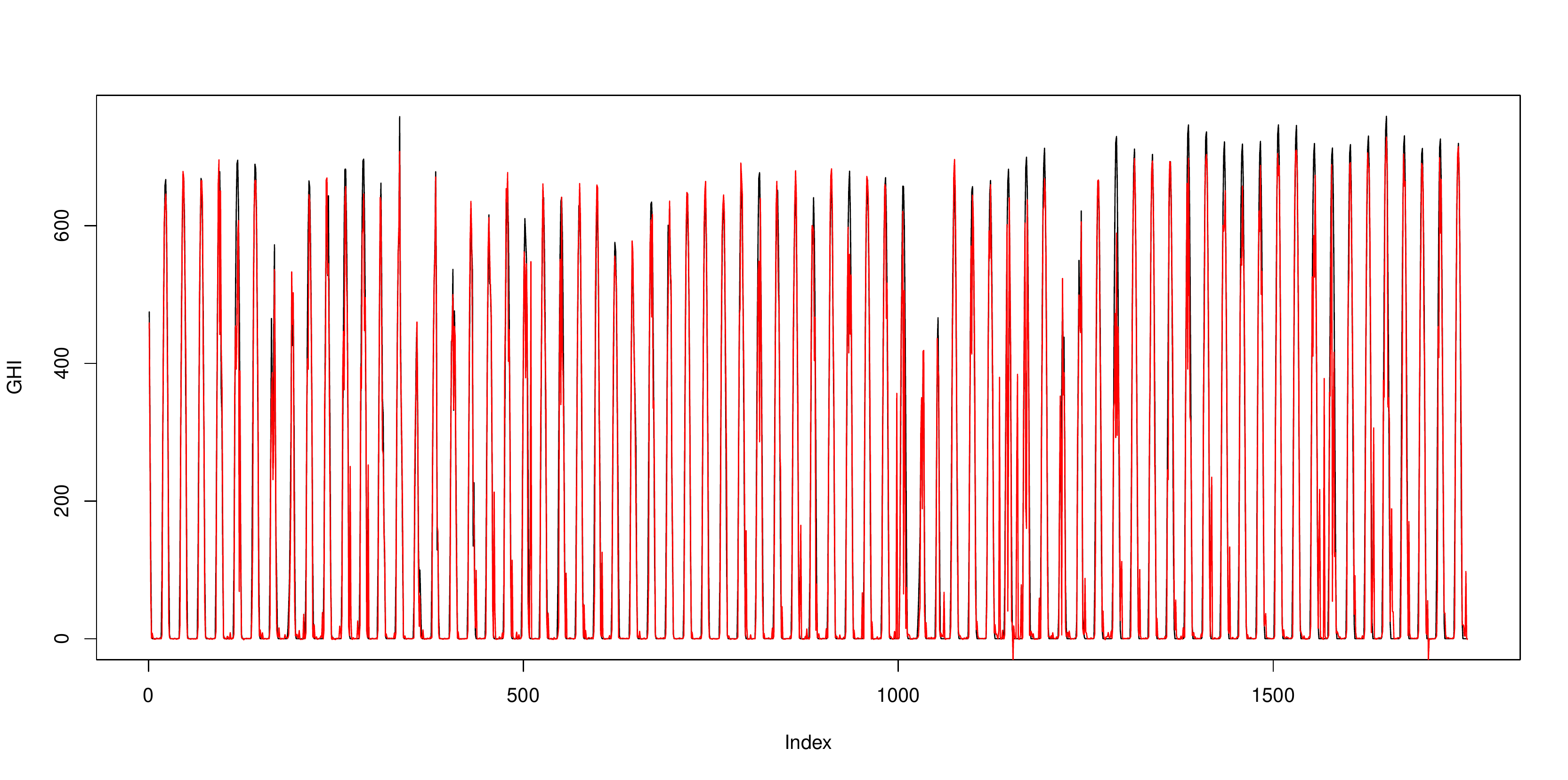}
		\caption{\bf GHI vs combined forecasts(using QRNN).}
		\label{fig:qrnn}
	\end{figure}

	\begin{table}[H]
		\centering
			\caption{Combined forecasts.}
			\label{tab:t6}
			\begin{tabular}{|l|l|l|} \hline
				Model & RSME & MAE \\ 		 \hline
				QRA   & 71.92 & 26.59 \\
				QRNN & 68.33 & 29.88\\
				PLAQR & 72.77 & 29.60\\
				OPERA & 78.53 & 35.51\\   \hline
			\end{tabular}
	\end{table}
	Table~\ref{tab:t6} show that the QRNN model has the most accurate forecasts based on evaluation metrics RMSE and MAE. GHI superimposed with combined predictions using QRNN for the testing set, 1759 observations which are approximately seventy-three days of forecasts, are shown in Fig. \ref{fig:qrnn}.
	
	\subsection{Murphy diagrams}
\begin{figure}[H]
		\centering
		\centering
		\includegraphics[width=1.0\linewidth, height=0.3\textheight]{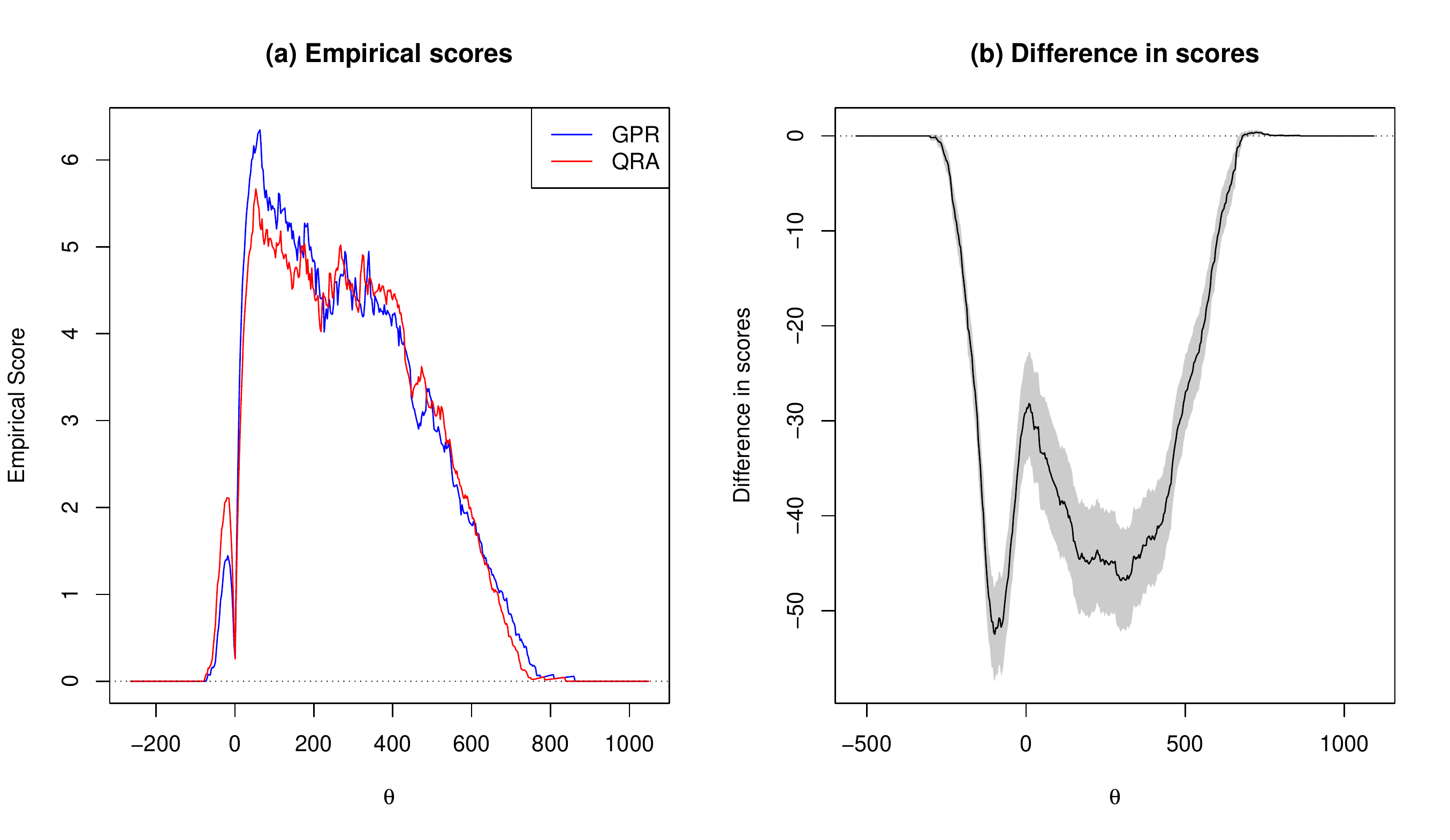}
		\caption{Murphy diagram for QRA vs GPR. {\bf Left panel:} (a) Empirical scores. {\bf Right panel:} (b) Difference in scores.}
		\label{fig:mur1}
	\end{figure}

\begin{figure}[H]
		\centering
		\centering
		\includegraphics[width=1.0\linewidth, height=0.3\textheight]{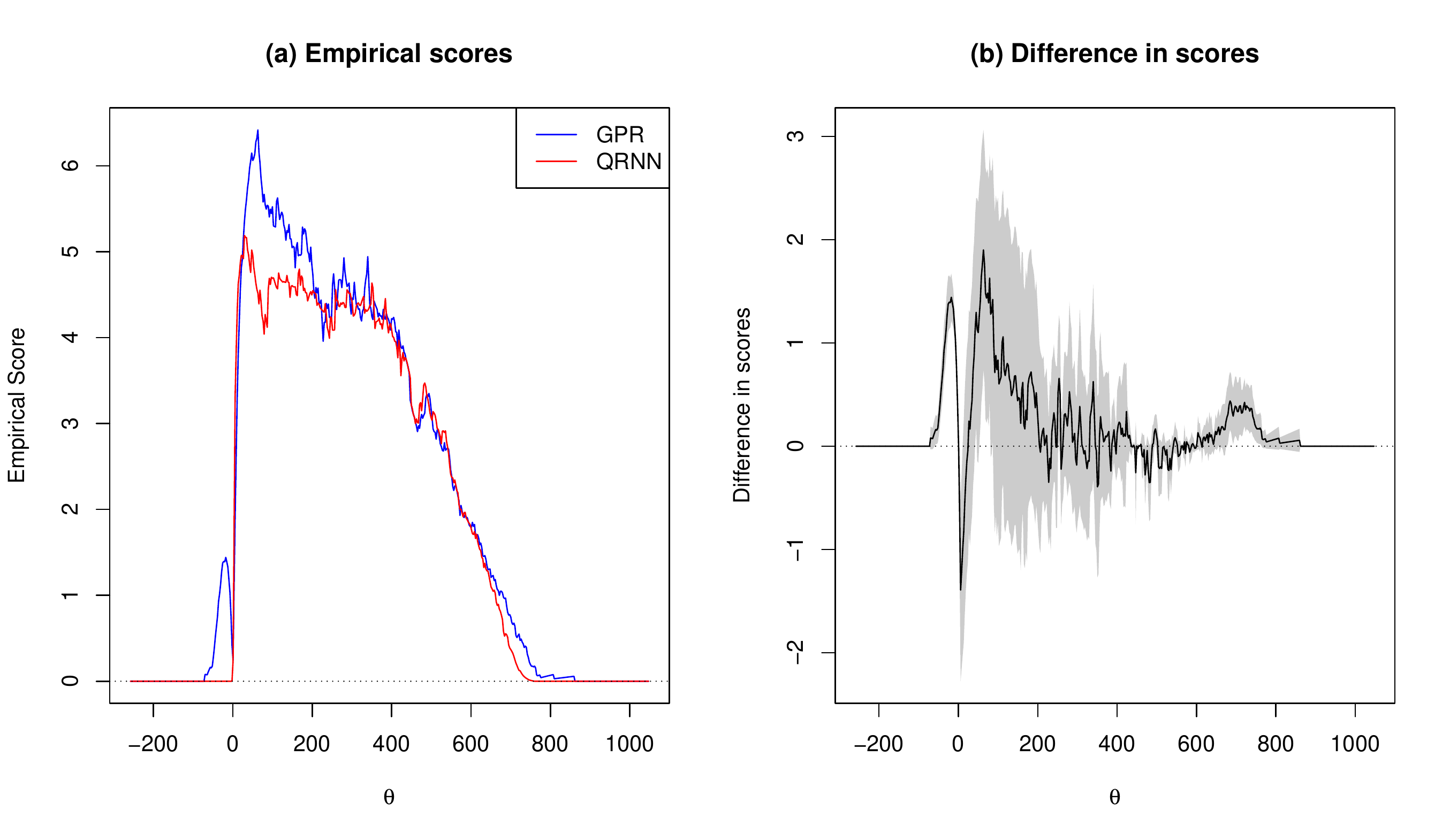}
		\caption{Murphy diagram for QRNN vs GPR.{\bf Left panel:} (a) Empirical scores. {\bf Right panel:} (b) Difference in scores.}
		\label{fig:mur2}
	\end{figure}

\begin{figure}[H]
		\centering
		\centering
		\includegraphics[width=1.0\linewidth, height=0.3\textheight]{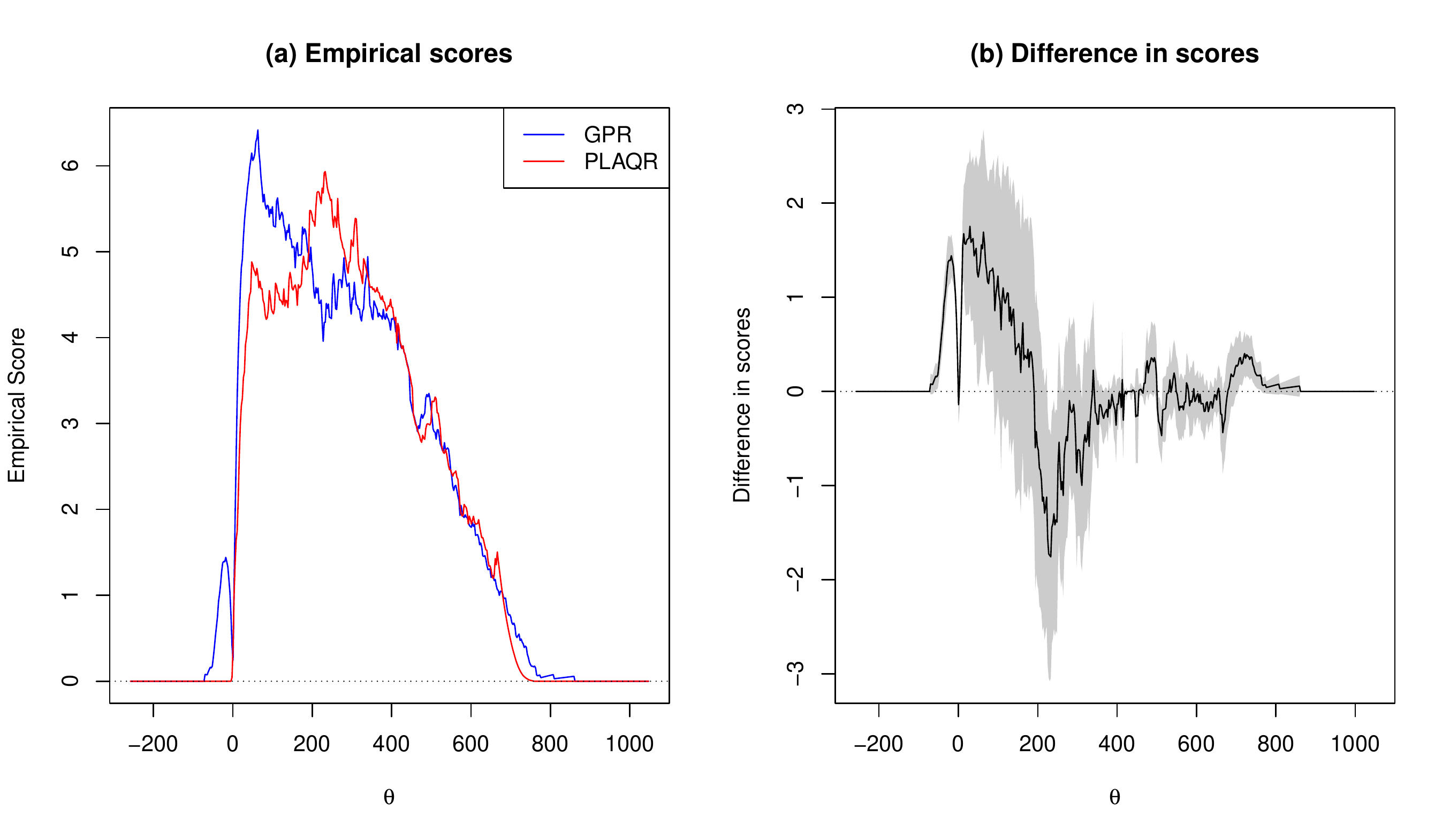}
		\caption{Murphy diagram for PLAQR vs GPR. {\bf Left panel:} (a) Empirical scores. {\bf Right panel:} (b) Difference in scores.}
		\label{fig:mur3}
	\end{figure}

\begin{figure}[H]
		\centering
		\centering
		\includegraphics[width=1.0\linewidth, height=0.3\textheight]{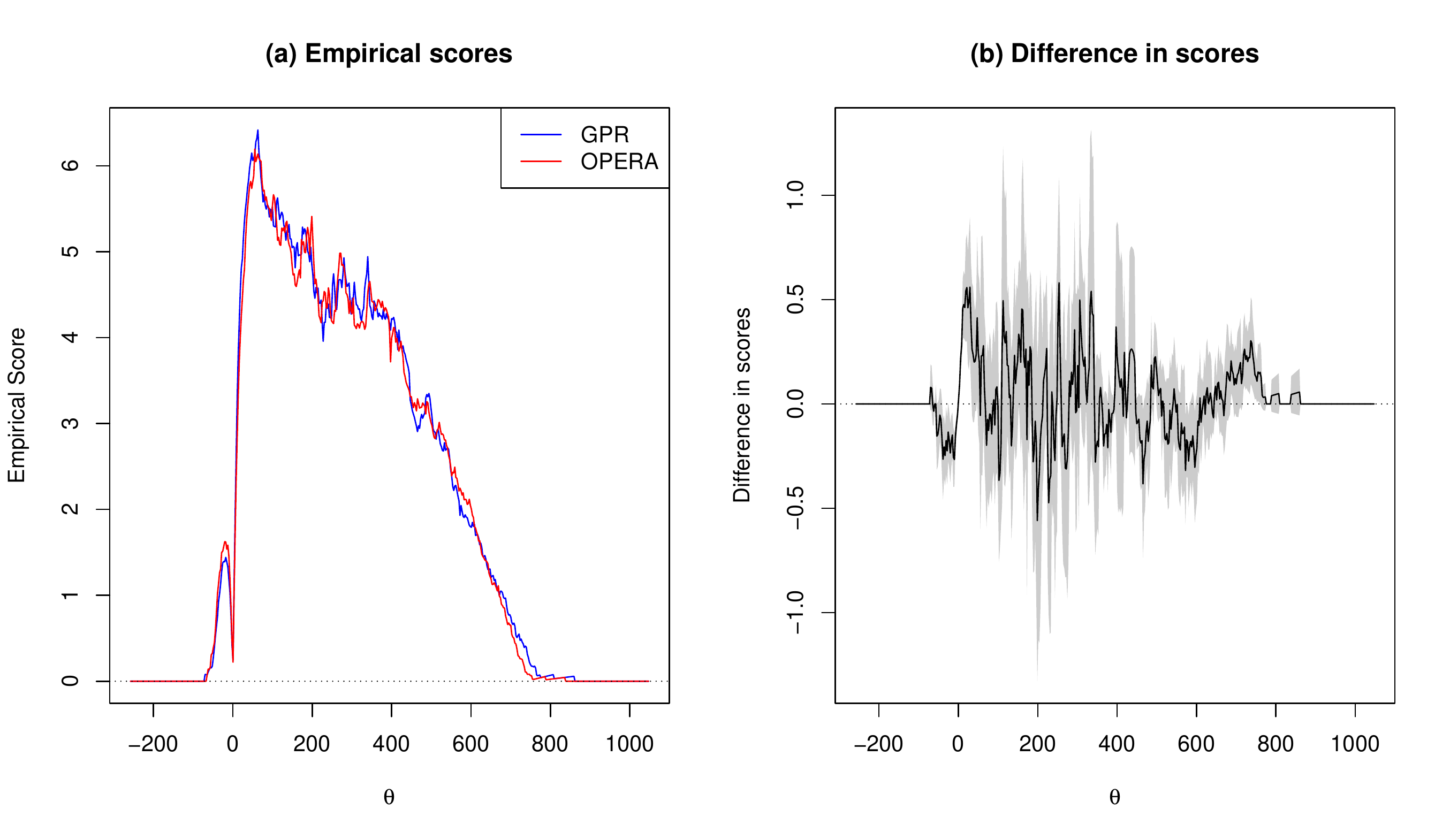}
		\caption{Murphy diagram for OPERA vs GPR. {\bf Left panel:} (a) Empirical scores. {\bf Right panel:} (b) Difference in scores.}
		\label{fig:mur4}
	\end{figure}

	The forecasting models were also evaluated using proper scoring rules, and Murphy diagrams were used to compare forecasts. Figures \ref{fig:mur1}~--~\ref{fig:mur4} show plots giving empirical scores and Differences in scores, and the GPR model proves superior to other models. The Murphy diagrams show that the Opera model is very good though the GPR model is a better forecaster than Opera. 

\begin{figure}[H]
		\includegraphics[width=0.9\linewidth, height=0.6\textheight]{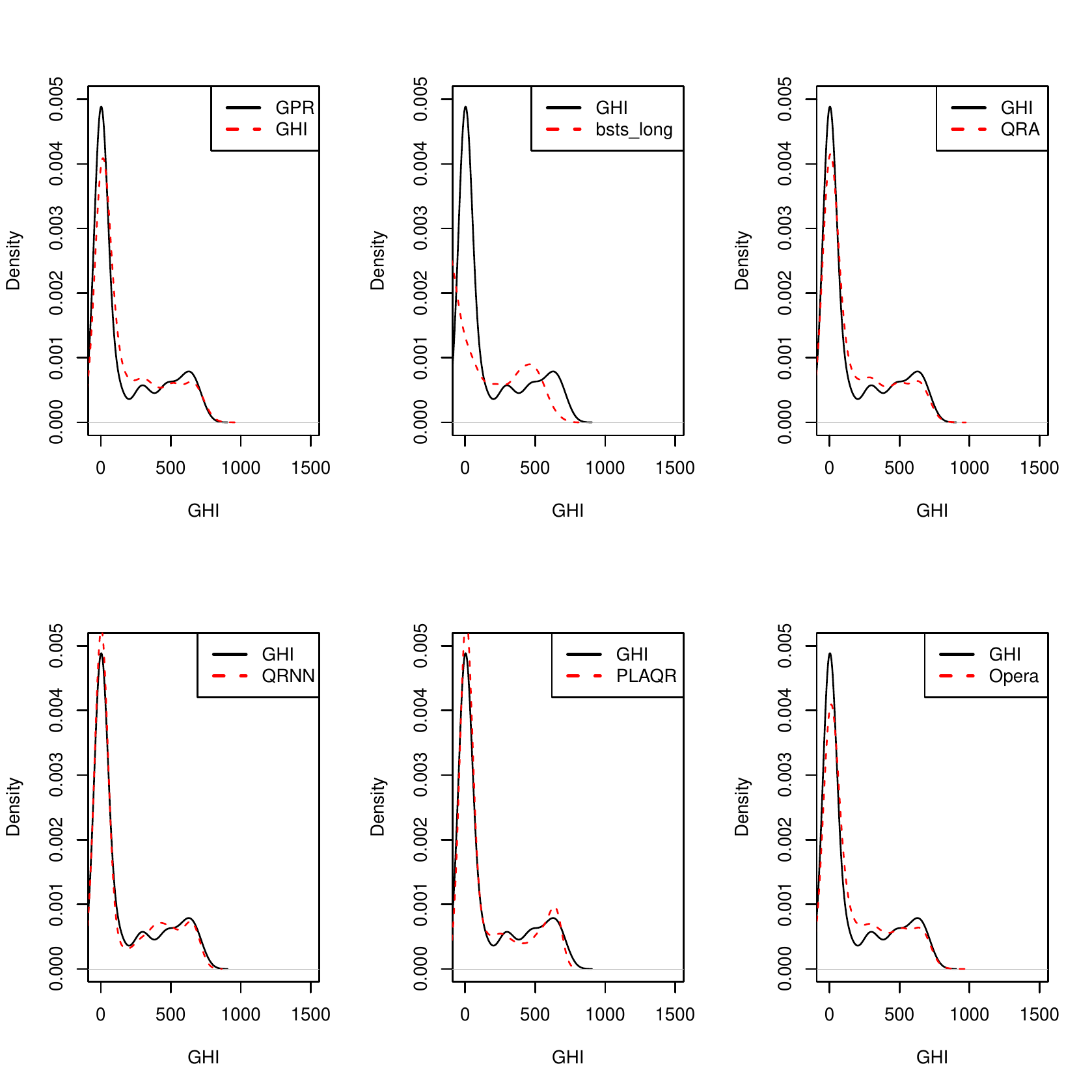}
		\caption{\bf Density plots.}
		\label{fig:models-density}
	\end{figure}

	Plots in Fig~\ref{fig:models-density} show a comparison between the densities of the models(in red) and GHI(in black). The plots show that models GPR, PLAQR, and QRNN are better forecasters. The densities of the forecasts are very close to the actual values of GHI.
	
	\subsection{Evaluating models using scoring rules}
	A further comparison of the models was made based on proper scoring rules. The proper scoring rules used are the logarithmic score (LogS), the continuous ranked probability score (CPRS), Dawid–Sebastiani Score (DSS), and Pinball Losses (PL). The forecasts which gave the lowest values of the scoring rules were the ones that gave a better forecasting performance.
	
	The scoring rules were computed as follows: we fit a parametric distribution to the forecasts. In this case, the Gamma distribution was the one that fitted the data and then evaluated the parameters. A comparison of the models is given in Table~\ref{tab:t7} for the models GPR, bstslong, QRA, QRNN, PLAQR, and Opera using probabilistic evaluations. The GPR model best fit based on the probabilistic evaluation metrics CRPS, LogS, DSS, and PL.
	\begin{table}[H]
		\centering
			\caption{Combined forecasts.}
			\label{tab:t7}
			\begin{tabular}{|l|l|l|l|l|} 		\hline
				Model & CRPS & LogS & DSS & PL \\ 			    \hline
				GPR  & 131.76 & 6.87 & 11.89 & 34.28 \\
				bstslong & 205.93 & Inf & 12.85 & 294.27 \\
				QRA   & 143.79 & Inf & 12.50 & 43.45 \\
				QRNN & 150.14 & Inf & 12.53 & 38.23\\
				PLAQR & 146.80 & Inf & 12.53 & 43.15\\
				OPERA & 142.15 & Inf & 12.48 & 37.40\\ 		\hline
			\end{tabular}
	\end{table}
	
	\section{Discussion and conclusion} \label{sec:Discussion}
	Decision-makers need robust models with high predictive capabilities in the electricity sector in medium-term forecasting of solar power, which is highly intermittent and has to be optimally integrated onto the grid. The GPR modelling framework, flexible in modelling nonstationary time series data and its ability to cope with abrupt regime changes, was found among the individual models to have the highest predictive ability in producing the most accurate and robust forecasts. The forecasts from the individual models were then combined using four combination methods, QRA, QRNN, PLAQR and OPERA. This increased the forecast accuracy, with the QRA being the best based on MAE and QRNN being the best based on the RMSE. Based on the evaluation measures for probabilistic forecasting, GPR has the lowest CRPS, DSS and PL scores proving to be the most superior individual model for predicting GHI. Using the Murphy diagrams, the GPR and QRNN models seem to have the same predictive abilities.
	
	Compared to previous findings, the present study reveals an upgrade to the work of Chandiwana ~\cite{Chandiwana}, Mpfumali ~\cite{Mpfumali}, among others from literature who did not include an ensemble of robust methods in medium-term forecasting of GHI. Using flexible and powerful techniques such as DGPR, GPR, AQR and BSTS, the present study produces robust medium-term forecasts of up to ten days ahead. A limitation of this study is using a two-layer DGPR model. Future research will explore the predictive abilities of DGPR models with more layers in the DGPR model.	
	
	This study explored the predictive abilities of robust models in the medium-term forecasting of GHI. A key finding of this result is that the GPR modelling framework provides robust predictions of medium-term solar power. This modelling framework provides nonlinear predictive capability with flexibility with nonstationary time series data. It can also cope with abrupt regime changes in the data and allows the inclusion of prior knowledge in the model specification. These results could be useful for system operators and decision-makers in power utility companies to integrate the intermittent renewable energy source into the grid.
	
	\section*{Acknowledgments}
	The authors are grateful to the numerous people for their helpful comments.

	%
	%
	
	%
	%
	%


\begin{thebibliography}{10}
		
		
		\bibitem{Yang}		
		D. Yang, C. Gu, Z. Dong, P. Jirutitijaroen, N. Chen and W.M. Walsh, Solar irradiance forecasting using spatial-temporal covariance
		structures and time-forward kriging, {\em Renewable Energy}, {\bf 60}, 235-245 (2013). \url{https://doi.org/10.1016/j.renene.2013.05.030}
		
		\bibitem{UNDP}
		UNDP, Sustainable Development Goal 7: Affordable and clean energy, [Online] Available  \url{https://www.undp.org/sustainable-development-goals#affordable-and-clean-energy}, Accessed on 4 April 2022.
		
		
		\bibitem{Brodersen}
		K.H. Brodersen, F. Gallusser, J. Koehlar, N. Remy and S.L. Scott, Infering causal impact using Bayesian structural time-series models, {\em The Annals of Applied Statistics}, {\bf 9(1)}, 247-274, (2015). \url{https://doi.org/10.1214/14-AOAS788}
		
		\bibitem{Bilionis}
		I. Bilionis, M.C. Emil and A. Mihai, Data-driven model for solar irradiation based on satellite observations, {\em Solar energy}, {\bf 110}, 22-38, (2014). \url{https://doi.org/10.1016/j.solener.2014.09.009}
		
		
		\bibitem{Tolba}
		H. Tolba, N. Dkhili, J. Nou, J. Eynard, S. Thil and S. Grieu, GHI forecasting using Gaussian process regression: Kernel study, {\em IFAC-PapersOnLine},
		{\bf 2(4)}, 455-460, (2019). \url{https://doi.org/10.1016/j.ifacol.2019.08.252}
		
		\bibitem{Wang}
		Y. Wang, F. Bo, H. Qing-Song and S. Li, Short-term solar power forecasting: A combined long short-term memory and gaussian process regression method,
		{\em Sustainability}, {\bf 13(7)}, 3665, (2021). \url{https://doi.org/10.3390/su13073665}
		
		\bibitem{Zhang}	
		Z. Zhang, W. Chao, P. Xiaosheng, Q. Hui, L. Hao, F. Jialong and W. Hongyu, Solar Radiation Intensity Probabilistic Forecasting Based on K-Means Time Series Clustering and Gaussian Process Regression, {\em IEEE Access}, {\bf 9}, 89079-89092, (2021).	\url{https://doi.org/10.1109/ACCESS.2021.3077475}
		
		\bibitem{Yu}
		B.M. Yu, J.P. Cunningham, G. Santhanam, S.I. Ryu, K.V. Shenoy and M. Sahani, Gaussian-process factor analysis for low-dimensional single-trial analysis of neural population activity, {\em Journal of neurophysiology}, {\bf 102(1)}, 614-635, (2009). \url{https://doi.org/10.1152/jn.90941.2008}
		
		
		\bibitem{Stonski}
		M. Stonski, Bayesian neural networks and Gaussian processes in identification of concrete properties, {\em Computer-Assisted Methods in Engineering and Science},  {\bf 18(4)}, 291-302, (2017). \url{https://cames.ippt.pan.pl/index.php/cames/article/view/108}
		
		
		\bibitem{Al-Shedivat}
		M. Al-Shedivat, A.G. Wilson, Y. Saatchi, Z. Hu and E.P. Xing,  Learning scalable deep kernels with recurrent structure, {\em The Journal of Machine Learning Research}, {\bf 18(1)}, 2850-2886, (2017). \url{https://dl.acm.org/doi/10.5555/3122009.3176826}
		
		\bibitem{Tsymbalov}
		E. Tsymbalov, S. Makarychev, A. Shapeev and M. Panov, Deeper connections between neural networks and Gaussian processes speed-up active learning, {\em Proceedings of the Twenty-Eighth International Joint Conference on Artificial Intelligence}, 3599-3605, (2019).	\url{https://doi.org/10.24963/ijcai.2019/499}
		
		\bibitem{Chandiwana}
		E. Chandiwana, C. Sigauke and A. Bere, Twenty-Four-Hour-Ahead Probabilistic Global Horizontal Irradiance Forecasting Using Gaussian Process Regression,
		{\em Algorithms}, {\bf 14(6)}, 177, (2021). \url{https://doi.org/10.3390/a14060177}
		
		\bibitem {Bates}
		J.M. Bates and C.W. Granger, The combination of forecasts, {\em Journal of the Operational Research Society}, {\bf 20(4)}, 451-468, (1969).
		\url{https://doi.org/10.1057/jors.1969.103}
		
		
		
		\bibitem{Saucer}
		A. Saucer, R.B. Gramacy and D. Hogdon, {\em Active Learning for Deep Gaussian Process Surrogates}, Available online: \url{https://arxiv.org/abs/2012.08015v2} (accessed on 25 November 2021).
		
		\bibitem{Radaideh}
		M.I. Radaideh, and T. Kozlowski, Surrogate modeling of advanced computer simulations using deep Gaussian processes, {\em Reliability Engineering \& System Safety}, {\bf 195}, 106731, (2020). \url{https://doi.org/10.1016/j.ress.2019.106731}
		
		\bibitem{Scott}
		S.L. Scott and H.R. Varian, Predicting the present with Bayesian structural time series, {\em Int. J. Math. Model. Numer. Optim}, {\bf 5(1-2)}, 4-23, (2014). \url{http://dx.doi.org/10.1504/IJMMNO.2014.059942}
		
		\bibitem{Mitchell}
		T.J. Mitchell and J.J. Beauchamp, Bayesian variable selection in linear regression, {\em Journal of the American statistical association}, {\bf 83(404)}, 1023-1032, (1988). \url{https://www.tandfonline.com/doi/abs/10.1080/01621459.1988.10478694}
		
		\bibitem{Madigan}	
		D. Madigan and A.E. Raftery, Model selection and accounting for model uncertainty in graphical models using Occam's window, {\em Journal of the American Statistical Association}, {\bf 89(428)}, 1535-1546, (1994). \url{https://doi.org/10.1080/01621459.1994.10476894}
		
		\bibitem{Harvey}
		A.C.Harvey, {\em Forecasting, structural time series models and Kalman filter}, Cambridge university press, (1990). 
		
		\bibitem{Durbin}
		J. Durbin and S.J. Koopman, A simple and efficient simulation smoother for state space time series analysis, {\em Biometrika}, {\bf 89(3)} 603-616, (2002).	
		\url{https://www.jstor.org/stable/4140605}
		
		Statistical Science
		1999, Vol. 14, No. 4, 382-417
		
		\bibitem{hoeting}
		J.A. Hoeting, D. Madigan, A.E Raftery and C.T. Volinsky, Bayesian model averaging: a tutorial, {\em Statistical Science}, {\bf 14(4)}, 382--401, (1999).
		\url{https://www.jstor.org/stable/pdf/2676803.pdf}
		
		\bibitem {gaillard2016}
		P. Gaillard, Y. Goude and R. Nedellec, Additive models and robust aggregation for GEFcom2014 probabilistic electric load and electricity price forecasting,
		{\em Int. J. Forecast}. {\bf 32(3)}, 1038-1050, (2016). \url{https://doi.org/10.1016/j.ijforecast.2015.12.001}
		
		\bibitem {fasiolo2020}
		M. Fasiolo, Y. Goude, R. Nedellec and S.N Wood, Fast Calibrated Additive Quantile Regression, {\em Journal of the American Statistical Association},
		{\bf 116(535)}, 1-12, (2020). \url{ https://doi.org/10.1080/01621459.2020.1725521}
		
		\bibitem{jordan}
		A. Jordan, F. Kr$\ddot{u}$ger and S. Lerch, Evaluating Probabilistic Forecasts with scoringRules, {\em Journal of Statistical Software}, {\bf 90(12)}, 1-37, (2019). \url{https://doi.org/10.18637/jss.v090.i12}
		
		\bibitem{Mpfumali}
		P. Mpfumali, C. Sigauke, A. Bere and S. Mulaudzi, Day ahead hourly global horizontal irradiance forecasting: An application to South African data,
		{\em Energies}, {\bf 12(18)}, 1-28, (2019). \url{https://doi.org/10.3390/en12183569}
	\end{thebibliography}
\end{document}